\documentclass{aastex}
\usepackage{graphicx}
\usepackage{color}
\usepackage{multirow}
\usepackage{amsmath}
\pdfoutput=1

\begin{document}
\title{The Origin of Parity Changes in the Solar Cycle}

\author{
Soumitra Hazra$^{1, 2}$\thanks{E-mail: soumitra.hazra@gmail.com}
and Dibyendu Nandy$^{1, 2}$ \thanks{E-mail:dnandi@iiserkol.ac.in}
\\
$^{1}$Department of Physical Sciences, Indian Institute of Science Education and Research Kolkata,  Mohanpur 741246, West Bengal, India\\
$^{2}$Center of Excellence and Space Sciences India, Indian Institute of Science Education and Research Kolkata, Mohanpur 741246, \\
West Bengal, India\\
}




\begin{abstract}
Although sunspots have been systematically observed on the Sun's surface over the last four centuries, their magnetic properties have been revealed and documented only since the early 1900s. Sunspots typically appear in pairs of opposite magnetic polarities which have a systematic orientation. This polarity orientation is opposite across the equator -- a trend that has persisted over the last century over which magnetic field observations exist. Taken together with the configuration of the global poloidal field of the Sun -- that governs the heliospheric open flux and cosmic ray flux at Earth -- this phenomena is consistent with the dipolar parity state of an underlying magnetohydrodynamic dynamo. Although, hemispheric asymmetry in the emergence of sunspots is observed in the Sun, a parity shift has never been observed. We simulate hemispheric asymmetry through introduction of random fluctuations in a computational dynamo model of the solar cycle and demonstrate that changes in parity are indeed possible in long-term simulations covering thousands of years. Quadrupolar modes are found to exist over significant fraction of the simulated time. In particular, we find that a parity shift in the underlying nature of the sunspot cycle is more likely to occur when sunspot activity dominates in any one hemisphere for a time which is significantly longer than the cycle period. We establish causal pathways connecting hemispheric asymmetry and cross-equatorial phase-shifts to parity flips in the underlying dynamo mechanism. Our findings indicate that the solar cycle may have resided in quadrupolar parity states in the distant past, and provides a possible pathway for predicting parity flips in the future.
\end{abstract}




\section{Introduction}

In 1843 Samuel Heinrich Schwabe identified the existence of the 11-year solar cycle in long-term sunspot observations which exist since the early 17th century. However, detailed observations regarding the nature of solar magnetic field exist only for the last hundred years \citep{hale19}.  Additionally, observations also reveal the systematic orientation associated with magnetic polarity of sunspots emerging on the solar surface. Sunspots, in general, appear in pairs with a leading and a following spot of opposite magnetic polarities. The magnetic polarity of the leading and following polarity spots belonging to different hemispheres are opposite, i.e.,  they are antisymetric across the equator. This can arise only from oppositely directed toroidal field belts in the two hemispheres of the Sun and is a manifestation of the dipolar nature of the underlying magnetic field. However, one may pose the question -- have the solar magnetic fields always been in a dipolar parity state? The limited span of solar magnetic field observations cannot address this question. 
 
To investigate this issue, we utilize a kinematic flux transport solar dynamo model which involves the generation and recycling of the toroidal and poloidal components of the solar magnetic field \citep{park55}. In this model, the toroidal field is produced by stretching of poloidal field lines at the base of the convection zone due to strong differential rotation \citep{park55} and the poloidal field is generated from the toroidal field through a combination of mean field $\alpha$-effect due to helical turbulence in the solar convection zone \citep{park55} and the Babcock-Leighton mechanism due to decay and dispersal of tilted bipolar sunspot region at the near-surface layers \citep{bab61,leig69}. The kinematic flux transport dynamo model based on the Babcock-Leighton mechanism for poloidal field generation has successful in explaining different observational aspects of the solar cycle \citep{dikp99,nand02,chat04, jouv07, goel09,nand11,dero12, pass14, hazr14, hazr16b}. Recent observations also lend strong support to the Babcock-Leighton mechanism as a primary source for poloidal field generation \citep{dasi10, muno13a}.

It is widely thought that stochastic fluctuations in the poloidal field generation mechanism is the primary source for irregularity in the solar cycle \citep{hoyn88, chou92,char20,char04}. In the Babcock-Leighton framework, poloidal field generation depends on the tilt angle of bipolar sunspot pairs, which is imparted by the action of Coriolis force on buoyantly rising toroidal flux tubes from the base of the solar convection zone. Observational scatter of tilt angles around the mean given by Joy's law may be produced by turbulent buffeting that a rising flux tube encounters during its journey through the convection zone \citep{long02}. Thus the Babcock-Leighton mechanism for poloidal field generation is not an entirely deterministic process but has inherent randomness \citep{chou07}. Another primary source in solar cycle irregularity is fluctuations in the meridional circulation \citep{lope09,kara10}.

On the one hand, two different types of symmetries are obtained, in general, in the solutions of the dynamo equations. The global magnetic field is of dipolar nature (dipolar or odd parity) if the toroidal field is antisymmetric across the equator (Fig.~1); Conversely, if the toroidal field is symmetric across the equator then the global field is of quadrupolar nature (quadrupolar or even parity; Fig.~1).  Some previous studies have found solutions that are of quadrupolar nature using low diffusivity in their kinematic dynamo models \citep{dikp01}. It has been suggested that an additional alpha effect at the base of the convection zone is essential to produce the observed dipolar parity \citep{dikp01, bona02}. However, other studies indicate that strong hemispheric coupling in presence of higher diffusivity is sufficient for generation of the global dipolar magnetic field without considering any additional alpha effect at the base of the convection zone \citep{chat04,chat06,hott10}. These past studies have been inspired with the primary aim of ensuring dipolar solutions to the dynamo equations with the notion that the solar dynamo has always persisted in the dipolar parity state with antisymmetric toroidal fields across the equator.

On the other hand, different solar activity in northern and southern hemispheres (known as hemispheric asymmetry) is well documented \citep{wald55, wald71, chow13, mccl13}. Observational evidence of strong hemispheric asymmetry exists during the onset of grand-minima like episodes \citep{soko94}. Theoretical and observational studies also suggest that hemispheric polar field at the minimum of the solar cycle can be used as a precursor to predict the amplitude of the next cycle \citep{scha78, scha05, jian07, kara12}. Thus the hemispheric asymmetry of polar field at solar minima may be responsible for the hemispheric asymmetry in the next cycle too. Utilizing kinematic solar dynamo models, some studies have been able to explain hemispheric asymmetry like phenomenon in the subsequent cycle by feeding the data of the polar flux of the previous cycles \citep{goel09}. Recent studies also demand that interplay between different dynamo modes may explain the origin of hemispheric asymmetry \citep{kapy16, shuk17, schu18}. Details about hemispheric coupling and hemispheric asymmetry can be found in a review paper by \cite{nort14}.

\begin{figure*}[!h]
\centering
\begin{tabular}{cc}
\includegraphics*[width=0.75\linewidth]{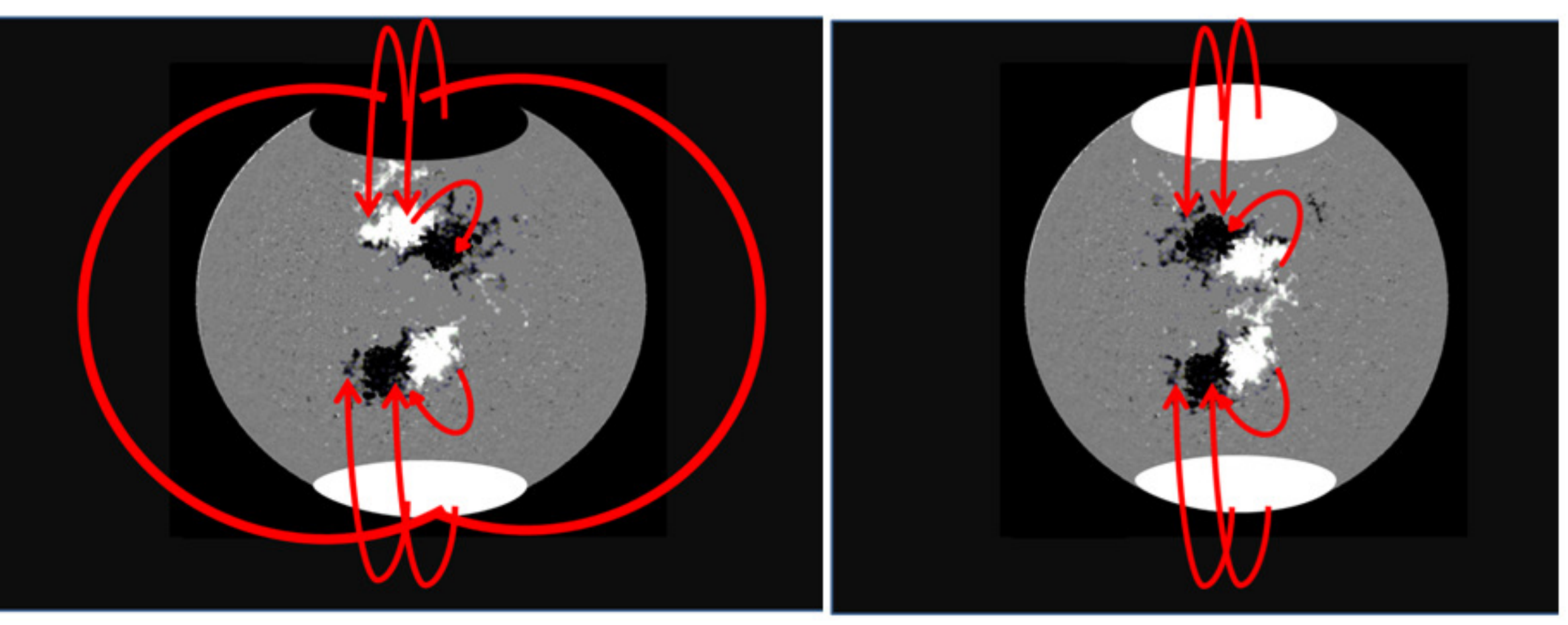} \\
\includegraphics*[width=1.0\linewidth]{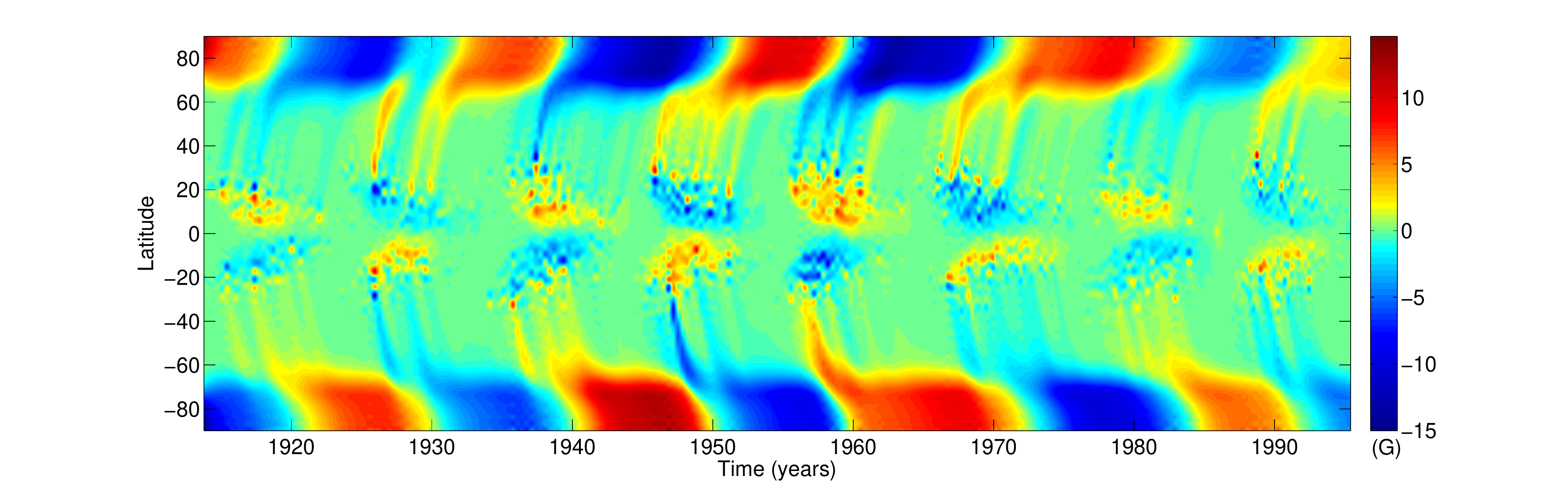} 
\end{tabular}                
\caption{\footnotesize{Top Panel shows dipolar (left) and quadrupolar (right) parity configuration. In the case of dipolar parity configuration (left image), polarity orientation is opposite across the equator; while in the case of quadrupolar parity configuration (right image), polarity orientation is same across the equator. Bottom panel shows the generated Butterfly diagram using observations fed into a surface flux transport model, indicating that the Sun has resided in a dipolar parity state as long as observations have existed.}}
\label{fig:1}
\end{figure*}

 To explore hemispheric asymmetry and parity issues and their inter-relationship, we first analyze the solar magnetic field in terms of axial dipolar and quadrupolar modes and find that parity reversal in the Sun may be related to hemispheric asymmetry. Then we try to verify these findings from our kinematic solar dynamo model. We introduce stochastic fluctuations in both the Babcock-Leighton mechanism and the additional mean field $\alpha$-effect and find dynamo solutions that can self-consistently change parity. The above result begs the question whether it is possible to predict parity flips in the Sun. We find that parity flips in the sunspot cycle tend to occur when solar activity in one hemisphere strongly dominates over the other hemisphere for a period significantly longer than the sunspot cycle timescale. However, strong domination of activity in one hemisphere does not necessarily guarantee a parity change.\\

\section{Model}
Our model is based on $\alpha\Omega$ dynamo equations in the axisymmetric spherical formulation wherein the dynamo equations are:
 
 \begin{equation}
   \frac{\partial A}{\partial t} + \frac{1}{s}\left[ \mathbf{v_p} \cdot \nabla (sA) \right] = \eta\left( \nabla^2 - \frac{1}{s^2}  \right)A + S (r, \theta, B)
\end{equation}

\begin{eqnarray}
   \frac{\partial B}{\partial t}  + s\left[ \mathbf{v_p} \cdot \nabla\left(\frac{B}{s} \right) \right] + (\nabla \cdot \mathbf{v_p})B = \eta\left( \nabla^2 - \frac{1}{s^2}  \right)B \nonumber \\ 
   + s\left(\left[ \nabla \times (A \bf \hat{e}_\phi) \right]\cdot \nabla \Omega\right)   + \frac{1}{s}\frac{\partial (sB)}{\partial r}\frac{\partial \eta}{\partial r},
\end{eqnarray}

where, $B (r, \theta)$ (i.e. $B_\phi$) and $A (r, \theta)$ are the toroidal component and vector potential for the poloidal component of the magnetic field respectively. Here $\Omega$ is the differential rotation, $\mathbf{v_p}$ is the meridional flow, $\eta$ is the turbulent magnetic diffusivity and $s = r\sin(\theta)$. For diffusivity and differential rotation profile, we use double step radial diffusivity profile ensuring a smooth transition to the low diffusivity beneath the base of the convection zone and an analytic fit of observed differential rotation. For meridional circulation, we use the same profile as described in \cite{hazr13}. In this present study, we use the parameters as given in \cite{hazr13} except we take $R_p =0.65 R_0$ (i.e. penetration depth of the meridional flow) and $v_0=17 m~s^{-1}$ (i.e. surface meridional flow speed).

In our model, toroidal field is generated due to strong differential rotation; while poloidal field is generated due to the combined effect of the Babcock-Leighton mechanism and the mean field alpha effect. In this paper, we model the Babcock-Leighton mechanism by the method of double ring \citep{durn97,hazr13, muno10}. In the double ring algorithm, we define the vector potential associated with each ring doublet as:
\begin{equation}
    A_{ar}(r,\theta, t)= K_1 A(\Phi, t)F(r)G(\theta),
\end{equation} 
where, $A(\Phi, t)$ defines the strength of ring doublet, and the constant $K_1$ ensures supercritical solutions. $\Phi$ is the magnetic flux. We use the profiles of F(r) and $G(\theta)$ as described in \cite{hazr13}. In this algorithm, we choose a latitude randomly in both northern and southern hemispheres and check whether the toroidal field strength at this latitude of the convection zone base exceeds the critical buoyancy threshold. If the toroidal field strength exceeds the buoyancy threshold then we remove a portion of the corresponding magnetic flux from this latitude at the base of the convection zone and place this flux at the surface in the form of ring doublets at the same latitude (see \cite{muno10,hazr13, hazr16a} for detailed explanation of the double ring algorithm).

We define the mean field $\alpha$-effect as:
 \begin{eqnarray}
    \alpha_{mf}= \alpha_{0mf} \frac{ \cos \theta }{4} \left[1+\textrm{erf}
    \left( \frac{r-r_1}{d_1}\right)\right]  
    \left[1-\textrm{erf}\left( \frac{r-r_2}{d_2}\right)\right] \nonumber \\ 
    \times \frac{1}{1+\left(\frac{B_\phi}{B_{up}}\right)^2} ~~~~
   \end{eqnarray}
where $\alpha_{0mf}$ controls the amplitude of this additional mean-field $\alpha$-effect, $r_1=0.71 R_\odot$, $r_2=R_\odot$, $d_1=d_2=0.25 R_\odot$, and $B_{up}= 10^4~G$ i.e. the upper threshold. The function $\frac{1}{1+\left(\frac{B_\phi}{B_{up}}\right)^2}$ ensures that this additional $\alpha$ effect is only effective on weak magnetic field strengths (below the upper threshold $B_{up}$) and the value of $r_1$ and $r_2$ confirms that this additional mechanism takes place inside the bulk of the convection zone.

We run our simulation without any fluctuation and find solar-like solution with always dipolar parity. 

\section{Results}
\subsection{Multipolar Expansions of Solar Magnetic Fields and Parity-Asymmetry Relationship}
We can express solar photospheric magnetic fields in terms of spherical harmonics. It can be written as,
   \begin{equation}
 B_r(\theta, \phi, t) = \sum_{l=0}^{l_{max}} \sum_{m=0}^{l} B_l^m (t) Y_l^m (\theta, \phi),
  \end{equation}
 where $\theta$ and $\phi$ are the colatitude and longitude respectively, and t is the time.
 The spherical harmonics $Y_l^m (\theta, \phi)$ are defined as,
 \begin{equation}
  Y_l^m (\theta, \phi) = (- 1)^m \sqrt{\frac{2l + 1}{4 \pi} \frac{(l - m)!}{(l + m)!}}~P_l^m (\cos \theta)~e^{im \phi},
  \end{equation}
 where $ P_l^m (\cos \theta)$ are the associated Legendre polynomials of degree $l$ and order $m$.
 Considering axial symmetry ($m=0$), we can write the expression of radial magnetic field in terms of axial dipolar and quadrupolar moments (assuming the axial dipolar and quadrupolar moments are the main determinants of radial magnetic field):
   \begin{equation}
 B_r(\theta, \phi, t) = C_1* DM* P_1(\cos(\theta)) + C_2* QM* P_2(\cos(\theta)),
  \end{equation}
where the dipolar (DM) and quadrupolar (QM) moments represent $B_1^0$ and $B_2^0$ respectively, $C_1= \sqrt{\frac{3}{4 \pi}}$ and $C_2= \sqrt{\frac{5}{4 \pi}}$. 
  Expression of radial magnetic field at a particular latitude for the northern hemisphere is given by,
\begin{equation}
 B_n= C_1* DM* \cos(\theta) + C_2* QM* \frac{1}{2} (3 \cos^2(\theta) -1),
  \end{equation}
 and for the southern hemisphere,
\begin{equation}
 B_s= - C_1* DM* \cos(\theta) + C_2* QM* \frac{1}{2} (3 \cos^2(\theta) -1).
  \end{equation} 
  Combining equations (8) and (9), we get,
 \begin{equation}
 DM = \frac{1}{2~C_1~\cos(\theta)} (B_n - B_s),
  \end{equation}
  and
\begin{equation}
 QM = \frac{1}{C_2~(3 \cos^2(\theta) -1)} (B_n + B_s).
  \end{equation}  
  Note that, here $B_n$ and $B_s$ are the signed magnetic field strengths in northern and southern hemispheres.\\
 So, for a particular latitude, we get,
 \begin{equation}
 \frac{QM}{DM} = C_3~\frac{B_n + B_s}{B_n - B_s} = C_3~\frac{\frac{B_n}{B_s} + 1}{\frac{B_n}{B_s} - 1},
  \end{equation}
   and
    \begin{equation}
 \frac{DM}{QM} = C_4~\frac{B_n - B_s}{B_n + B_s} = C_4~\frac{\frac{B_n}{B_s} - 1}{\frac{B_n}{B_s} + 1},
  \end{equation}
  
where $C_3$ and $C_4$ are constants for a particular latitude. We assume these constants to be equal to unity for simplicity of calculation. We find that relative strengths of signed axial quadrupolar (QM) and dipolar (DM) moments depend on the ratio of signed magnetic field strengths between northern and southern hemispheres i.e., $\frac{B_n}{B_s}$.  
\begin{figure}
\includegraphics*[width=0.9\linewidth]{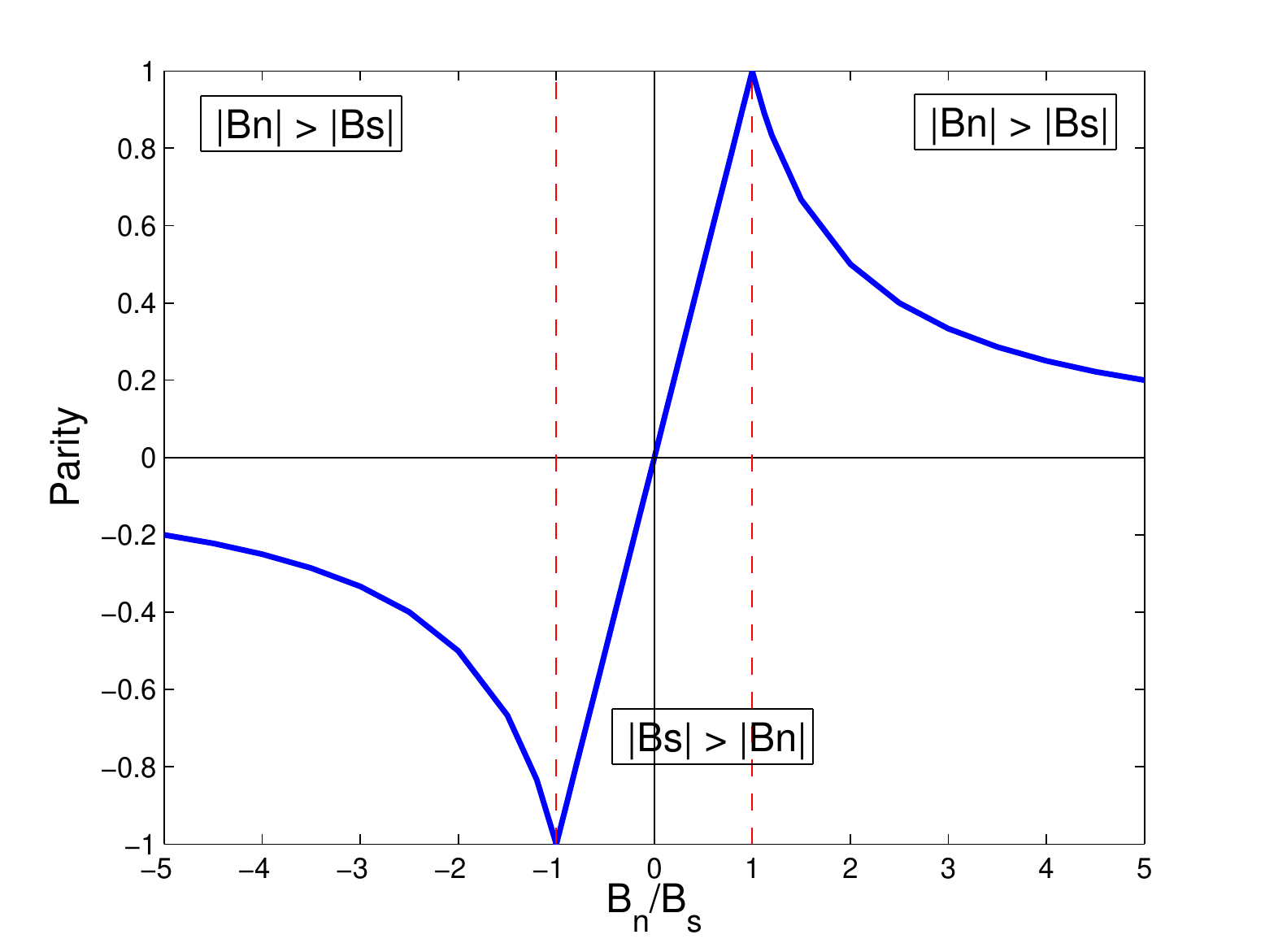}  
\caption{\footnotesize Dependence of parity function P(t) on $\frac{B_n}{B_s}$. Region between two dashed line indicates the region where the unsigned magnetic field strengths in the southern hemisphere are greater than the northern hemisphere.}
\label{fig:2}
\end{figure}
 As parity is essentially the measure of relative strength between quadrupolar and dipolar modes of solar magnetic fields, we can also define parity function $P(t)$ in terms of quadrupolar and dipolar moments.\\
\begin{eqnarray} 
 P(t) & = & \frac{|QM| - |DM|}{|QM| + |DM|}, \\
 & = & \left\{\begin{array}{cc}  \frac{|\frac{QM}{DM}| - 1}{|\frac{QM}{DM}| + 1}  ~~~~~~~~~~~~~~~& \frac{QM}{DM} \geq 0\\\\
       \frac{1- |\frac{DM}{QM}|}{1 + |\frac{DM}{QM}|}         ~~~~~~~~~~~~~~~ & \frac{QM}{DM} < 0
    \end{array}\right.,
  \end{eqnarray}
 where QM is the quadrupolar moment, and DM is the dipolar moment. Value of parity function is -1 for dipolar parity and +1 for quadrupolar parity.\\

Fig.~2 shows the dependence of parity function P(t) on the ratio of signed magnetic field strength between northern and southern hemispheres. From Fig.~2, we find that parity shift is associated with the change in the relative absolute magnetic field strengths between two hemispheres (i.e., $|B_n|/|B_s|$). This result indicates that nonlinear coupling between dipolar and quadrupolar modes of solar magnetic fields across the hemisphere may be an important factor in characterizing parity reversals. Some recent studies also indicate that coupling between different modes are responsible for long term solar variability as well as hemispheric asymmetry \citep{kapy16, shuk17, schu18}.

 \subsection{Parity-Asymmetry Relationship from Mean-field Kinematic Solar Dynamo Model}
 \begin{figure*}
        \centering
       \includegraphics[height=3.5cm, width=15 cm]{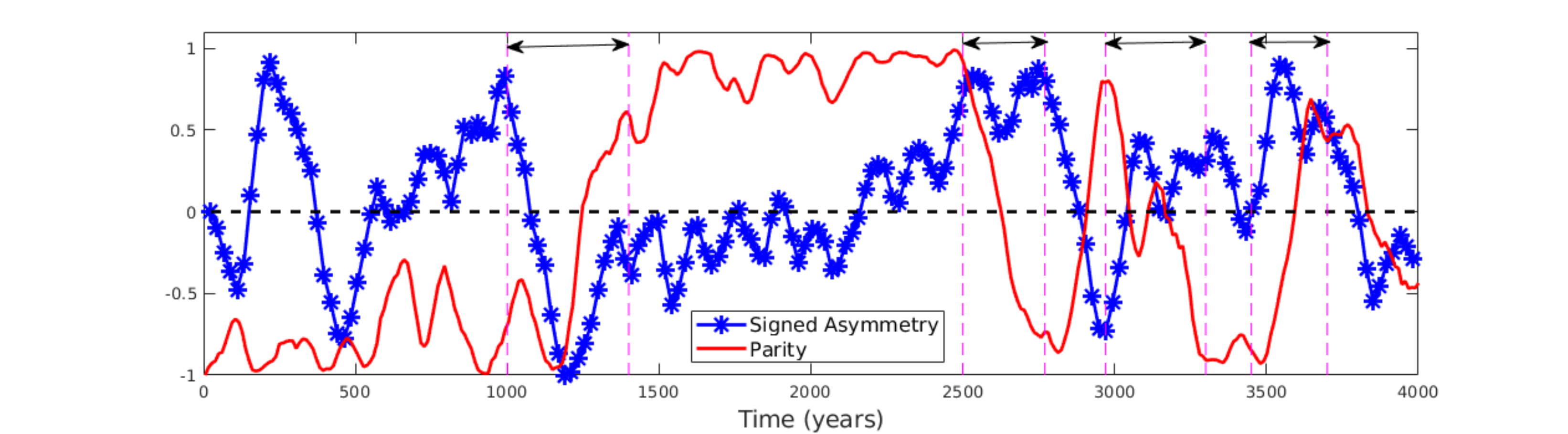}
        \includegraphics[height=3cm, width=15 cm]{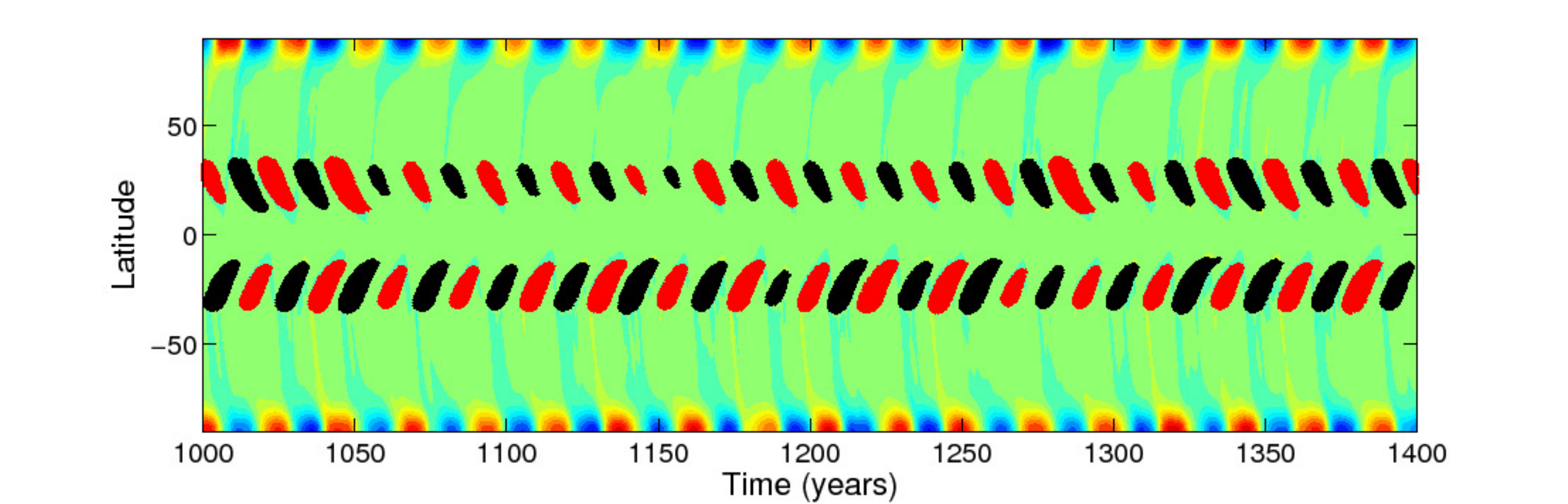}
        \includegraphics[height=3cm, width=15 cm]{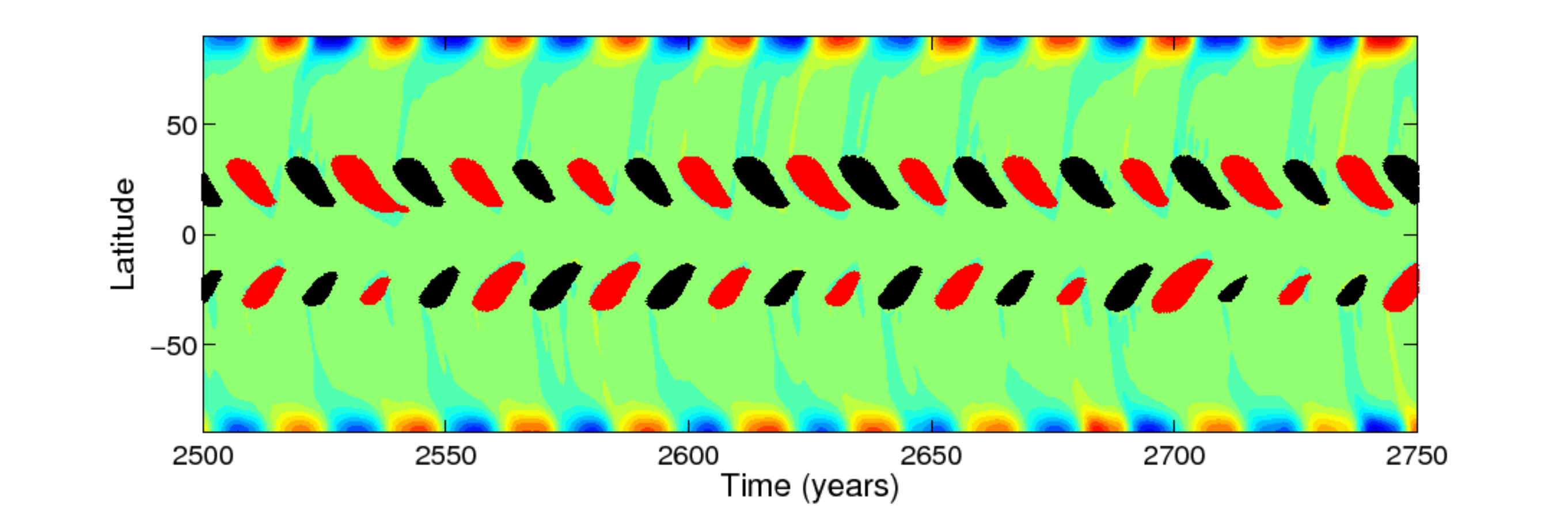}
        \includegraphics[height=3cm, width=15 cm]{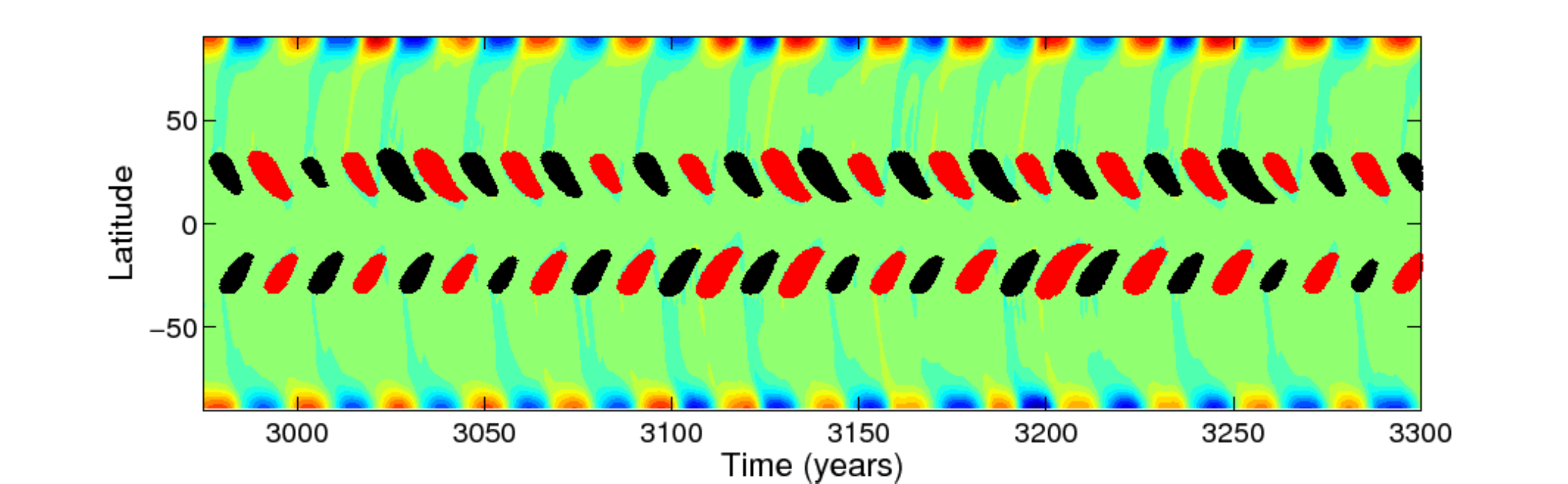}
        \includegraphics[height=3cm, width=15 cm]{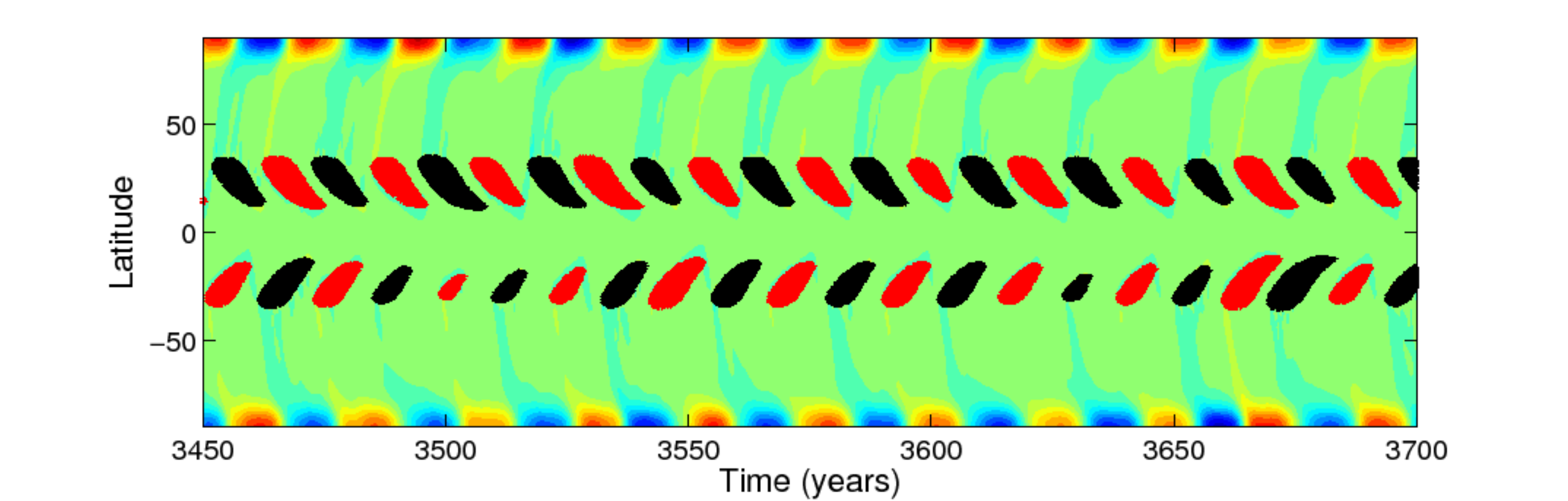}
        \caption{\footnotesize First panel shows the evolution of smoothed parity (red colour) and 22 year averaged smoothed normalized signed asymmetry (blue color) obtained from our simulations. Second, third, fourth and fifth panels are the simulated butterfly diagrams for different time intervals where parity change takes place. Selected time intervals are shown in top panel by double arrow. All these plots indicate that a change in solar parity takes place when sunspot activity in one hemisphere dominates over the other for a sufficiently large period of time. This simulations corresponds to 60\% fluctuations in Babcock-Leighton mechanism and 50 \% fluctuations in mean field $\alpha$.}
        \label{fig:3}
\end{figure*}

\begin{figure*}
        \centering
        \includegraphics[height=3.5cm, width=15 cm]{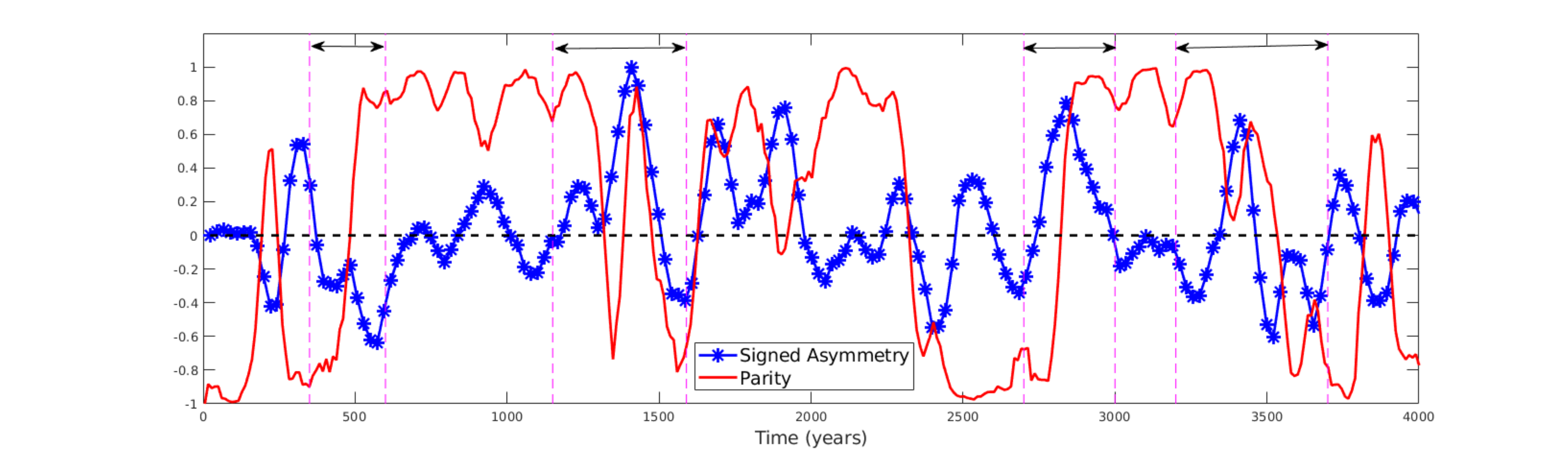}
        \includegraphics[height=3cm, width=15 cm]{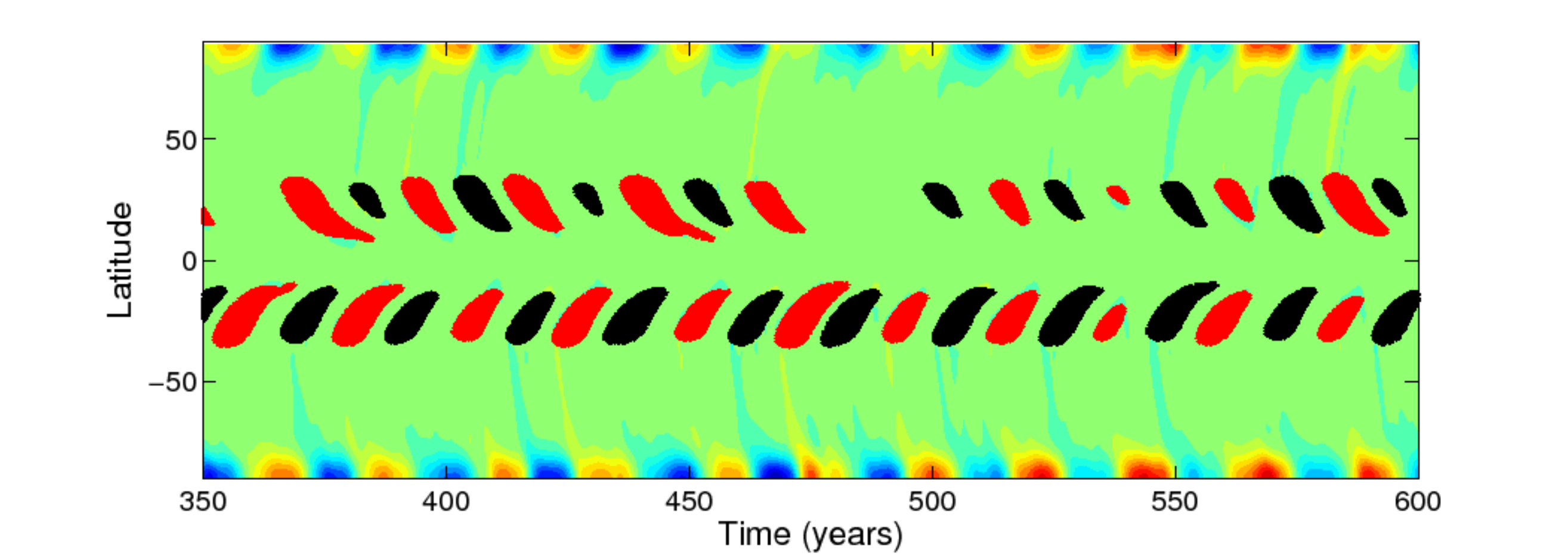}
        \includegraphics[height=3cm, width=15 cm]{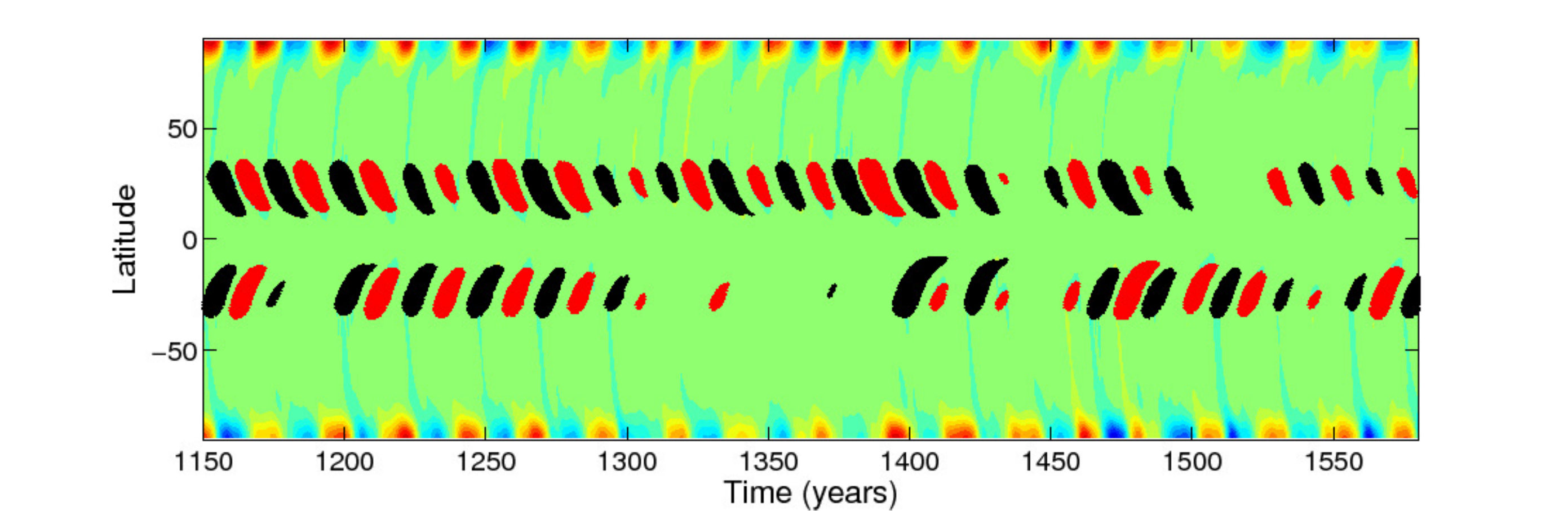}
        \includegraphics[height=3cm,width=15 cm]{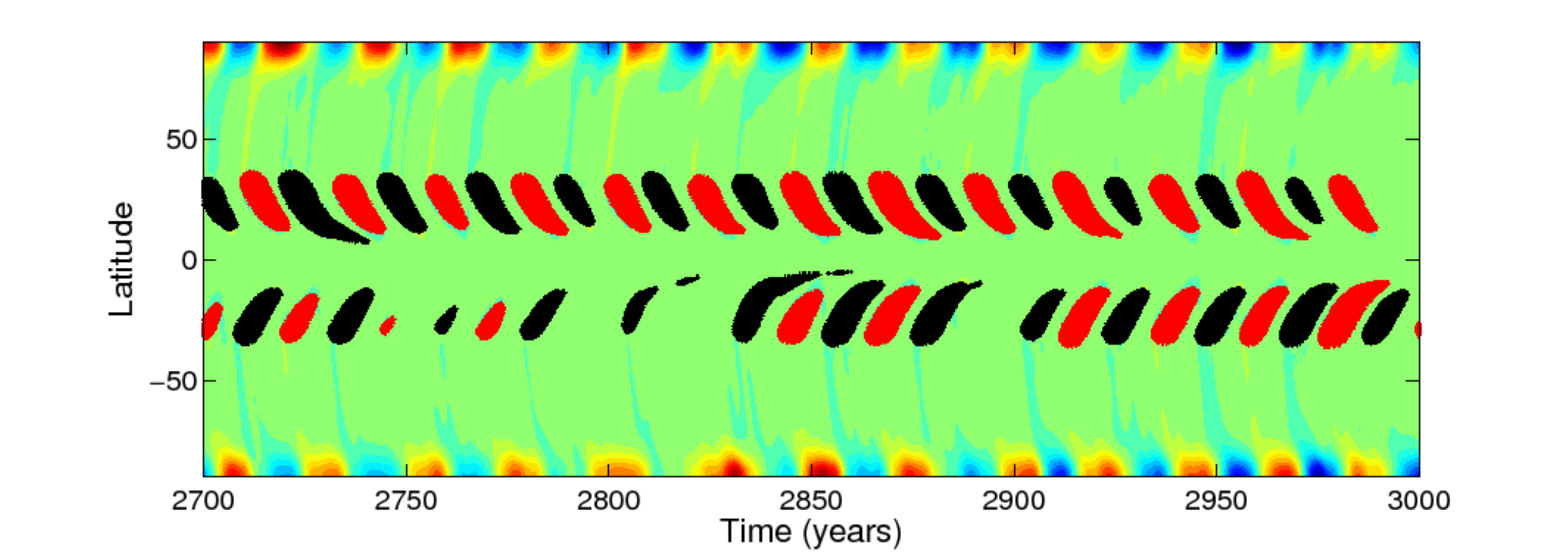}
        \includegraphics[height=3cm,width=15 cm]{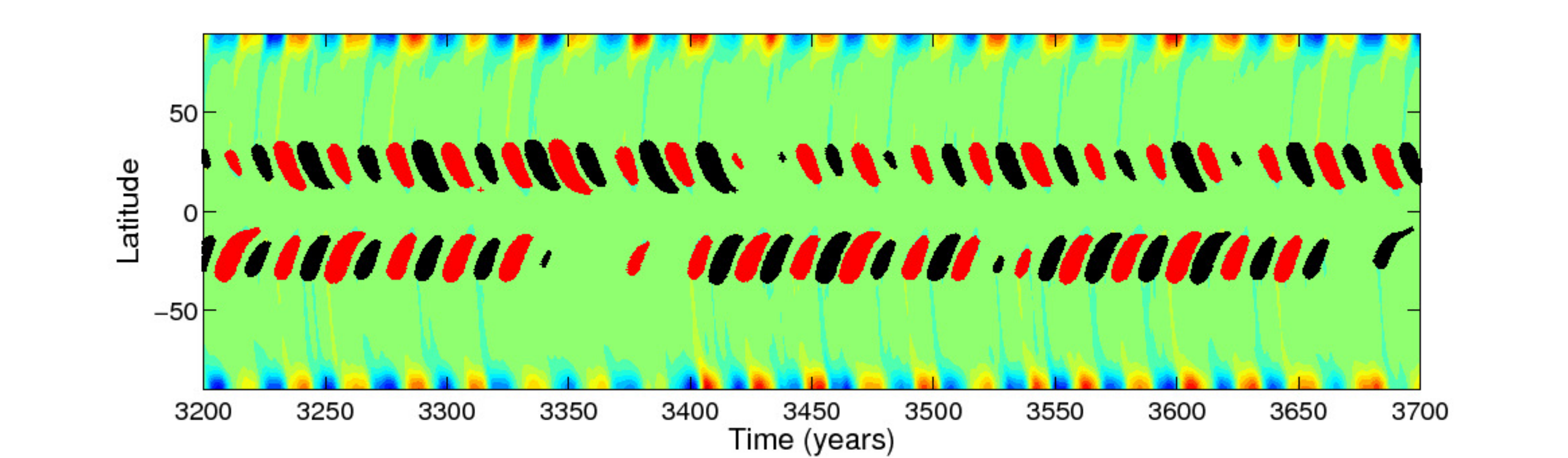}
     \caption{\footnotesize First panel shows the evolution of smoothed parity (red colour) and 22 year averaged smoothed normalized signed asymmetry (blue color) obtained from our simulations. Second, third, fourth and fifth panels are the simulated butterfly diagram for different time intervals where parity change takes place. Selected time intervals are shown in top panel by double arrow.  These simulations indicate solar cycle parity changes take place when sunspot activity in one hemisphere dominates over the other for a sufficiently large period of time. This simulation corresponds to 75\% fluctuation in Babcock-Leighton mechanism and 150 \% fluctuation in mean field $\alpha$.}
        \label{fig:4}
\end{figure*} 

We investigate our theoretical findings about parity-asymmetry relationship in detail using the kinematic Babcock-Leighton solar dynamo model. To explore the parity issue with our kinematic dynamo simulations, we calculate the parity function in terms of quadrupolar and dipolar moments following Eq. 15. We calculate quadrupolar and dipolar moments following Eq. 12 and 13 in terms of $B_n$ and $B_s$:

  \begin{equation}
 B_n=  \int_{t-T/2}^{t+T/2} B_N(t')dt'
 \end{equation}
  \begin{equation}
 B_s=  \int_{t-T/2}^{t+T/2} B_S(t')dt'
 \end{equation}
 where $B_N$ and $B_S$ are the amplitudes of the toroidal field at $25^{\circ}$ latitude in both northern and southern hemispheres at the base of the solar convection zone, and T is the cycle period in any one hemisphere. Value of the constants $C_3$ and $C_4$ (appeared in Eq. 12 and 13) are calculated for $25^{\circ}$ latitude. The value of parity function should be +1 for quadrupolar parity and -1 for dipolar parity. In the first scenario, we run dynamo simulations without fluctuations, considering both the Babcock-Leighton mechanism and mean field alpha effect as a poloidal field generation process. We find the parity of the solutions are always dipolar.
 
 Earlier studies have indicated that the dipolar parity of dynamo solutions is associated with strong hemispheric coupling -- which can be obtained either by increasing diffusivity \citep{chat04} or by introducing an additional mean field $\alpha$ effect (distributed through the convection zone, or tachocline; \cite{dikp01}). However, none of these models consider stochastic fluctuation in their simulations. In reality, the Babcock-Leighton mechanism is not a fully deterministic process but has some intrinsic randomness. This random nature arises due to scatter in tilt angles (an observed fact) of bipolar sunspot pairs whose underlying flux tubes are subject to turbulent buffeting during their ascent through the turbulent convection zone \citep{long02}. The other poloidal field generation mechanism, namely mean field alpha effect, is also inherently random as this mechanism arises due to helical turbulence inside the convection zone. Motivated by this fact, we introduce stochastic fluctuations in both the Babcock-Leighton source ($K_{ar}$) and mean field poloidal source terms ($\alpha_{mf}$). We find that parity of dynamo solutions oscillate between dipolar and quadrapolar modes (see top panels of Fig.~3 and 4).
 
 What is the cause of parity change in our model? One possible reason is the different levels of fluctuations in poloidal field source terms associated with northern and southern hemispheres. Stochastic fluctuations or randomness in the poloidal source is plausibly at the heart of hemispheric asymmetry \citep{hoyn88}. We find no north-south asymmetry in the simulated solar cycle by performing dynamo simulations without stochastic fluctuation. Thus we speculate there is a relationship between hemispheric asymmetry and parity change. To investigate the relationship between parity and hemispheric asymmetry, we need to define hemispheric asymmetry in the context of our simulations. In our kinematic flux transport dynamo model, we model the Babcock-Leighton mechanism by the double ring algorithm. We believe that this algorithm to be a more realistic way to capture the essence of the Babcock-Leighton mechanism as well as sunspots. For this work, we take the difference between number of double ring eruptions in the northern and southern hemispheres as a measure of hemispheric asymmetry. We define this difference the signed asymmetry for the rest of the paper.
 
 \begin{figure*}
 \centering
\includegraphics[width=15 cm]{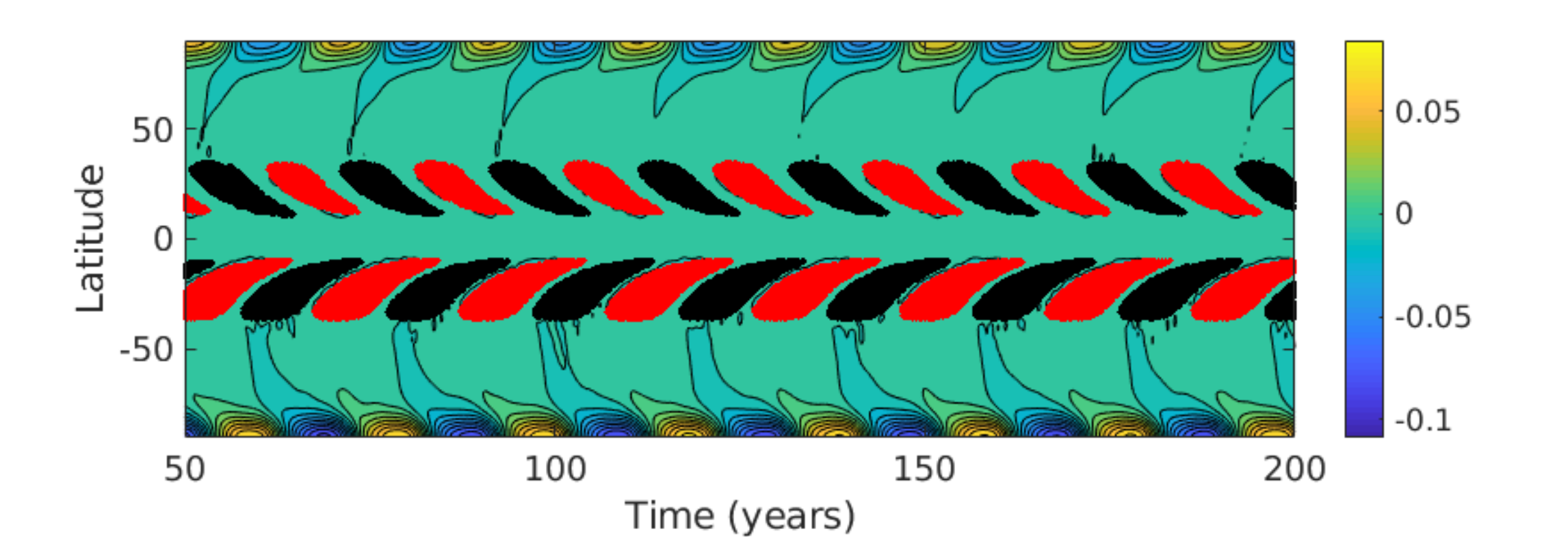} 
\includegraphics[width= 15 cm]{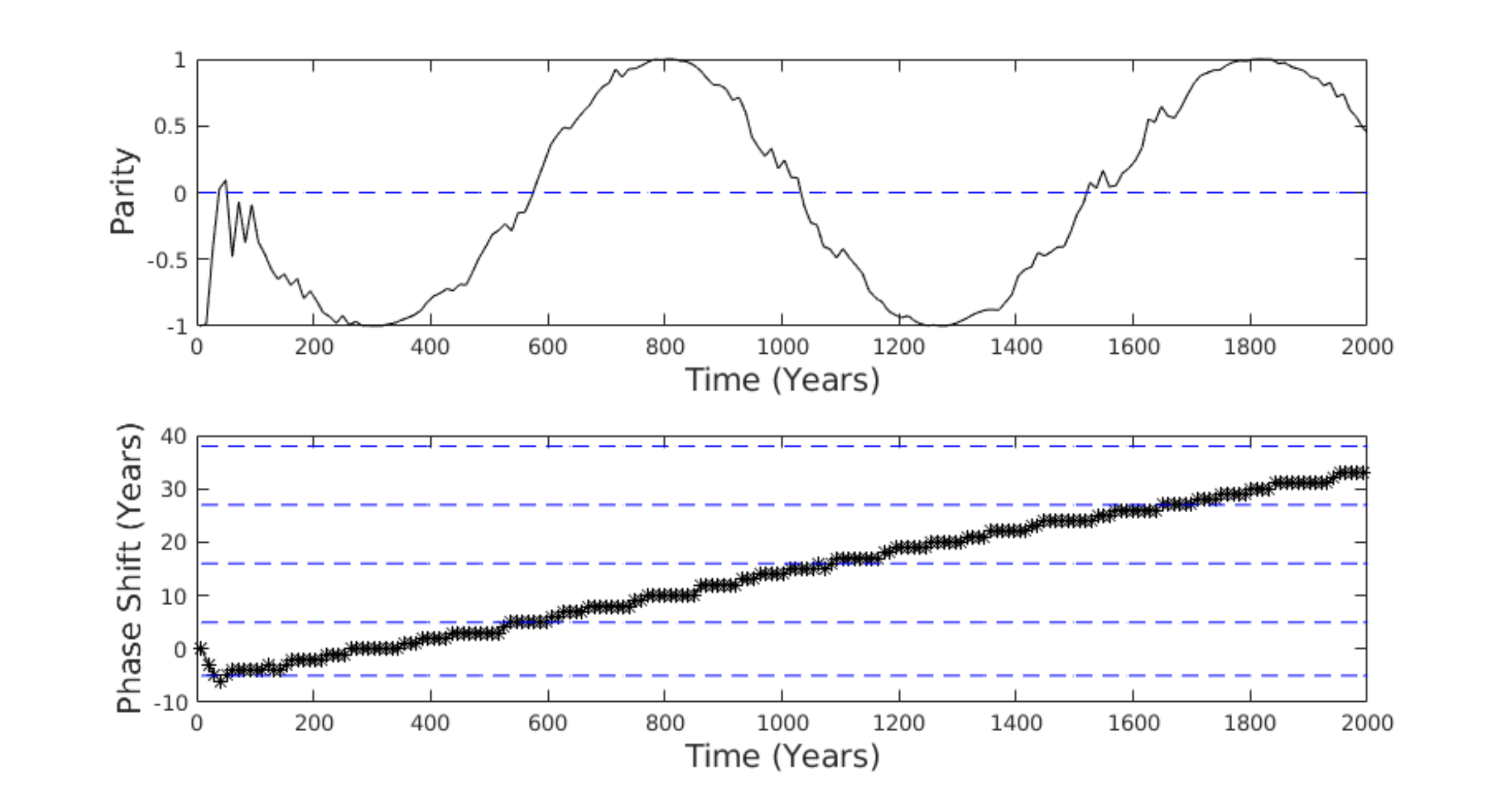}
\caption{Top panel shows the simulated buttefly diagram when southern hemisphere is stronger than northern hemisphere throughout the simulation. Middle panel shows the evolution of parity which indicates that parity is oscillating between dipolar and quadrupolar mode with regular interval. The bottom panel shows how much cyclic activity in the northern hemisphere shifts over the southern hemisphere with time (phase shift variation) with time (i.e., $T^i_N-T^i_S$ where $T^i_N$ and $T^i_S$ is the time of $i^{th}$ cycle minima in the northern and southern hemispheres respectively). The blue dotted line indicates the phase shift duration when parity change will take place (-5, 5, 16, 27 years). This result confirms our previous result that the Sun is more likely to shift the parity when one hemisphere is sufficiently stronger over the other for few cycles.}
\label{fig:5}
  \end{figure*}
  
\begin{figure*}
        \centering
        \includegraphics[width=15 cm]{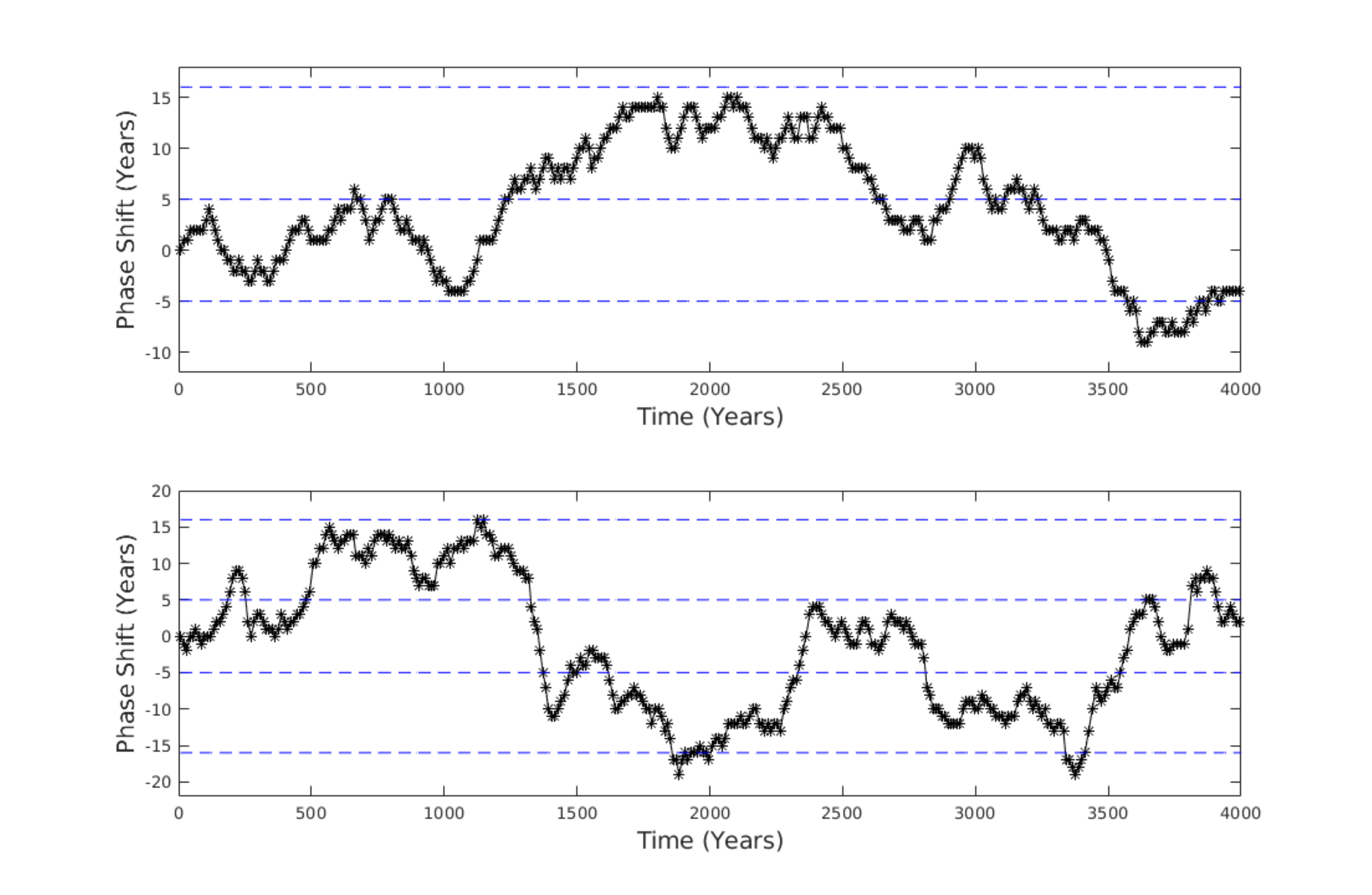}
        \caption{\footnotesize Both top and bottom panel shows how much cyclic magnetic activity in northern hemisphere progresses over the southern hemisphere with time (phase shift variation) with time. We define the phase shift as a time by which cyclic magnetic activity in northern hemisphere shifts over the southern hemisphere due to hemispheric asymmetry (i.e., $T^i_N-T^i_S$ where $T^i_N$ and $T^i_S$ is the time of  $i^{th}$ cycle minima in the northern and southern hemispheres respectively). The blue dotted line indicates the phase shift duration when parity change will take place (-5.5, 5.5, 16.5 years). Top panel corresponds to the result of simulation with 60\% fluctuations in Babcock-Leighton mechanism and 50 \% fluctuations in mean field $\alpha$; while bottom panel corresponds to the result of simulation with 75 \% fluctuation in Babcock-Leighton mechanism and 100 \% fluctuation in mean field $\alpha$.}
        \label{fig:6}
\end{figure*}

\begin{figure*}
\centering
\includegraphics[width=15 cm]{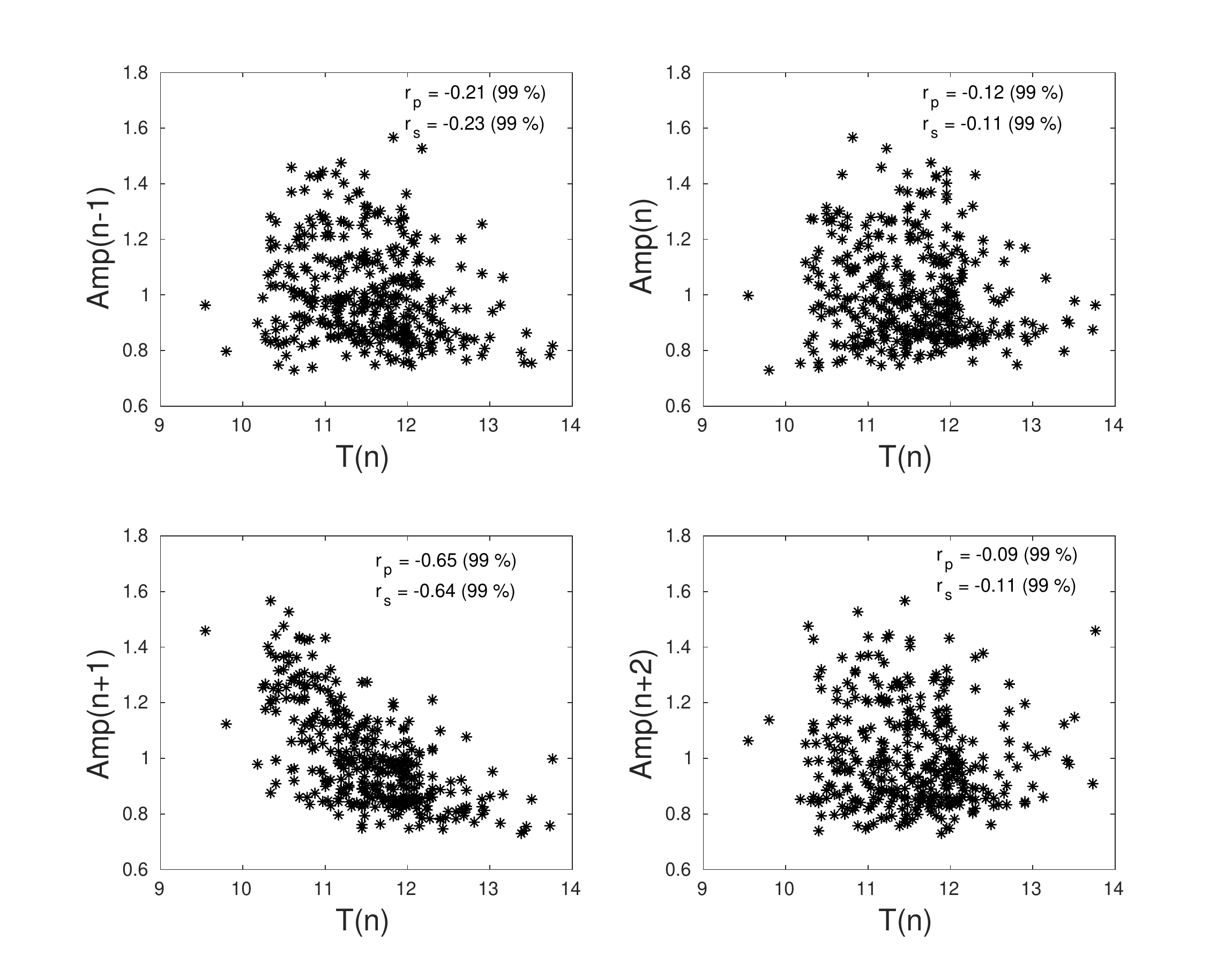}
\caption{\footnotesize Cycle-to-cycle correlations, between cycle period, T(n) and (a) cycle amplitude Amp(n-1) (b) Amp(n) (c) Amp(n+1) (d) Amp(n+2). The Pearson ($r_p$) and Spearman ($r_s$) correlation coefficients along with significance levels are inscribed.}
        \label{fig:7}
\end{figure*}  

 Figures 3 and 4 is the representative plot of parity and signed asymmetry relationship with the different level of fluctuations in the Babcock-Leighton mechanism and mean field $\alpha$-effect. The top panel in Fig.~3 and 4 shows the time evolution of parity and 22 years averaged signed asymmetry. A comparison between the time evolutions of parity and smoothed signed asymmetry reveals that change in parity is associated with the strong dominance of emergances in one hemisphere for a long period. Same phenomenon is reflected in the simulated butterfly diagrams. The second, third, fourth and fifth panels of Fig.~3 and 4 are the corresponding butterfly diagrams for different time intervals where parity change takes place. However, we also notice that on some rare occasions there is a strong dominance of eruptions in one hemisphere, but the parity does not change. Theoretical calculations also suggest that parity shifting is related to the relative magnetic field strength between two hemispheres. We have performed some additional simulations where one hemisphere strongly dominates over other throughout the simulation, to further validate our results. In this scenario, we always find the oscillation of parity with a regular interval (dipolar to quadrupolar and vice versa); which put our findings based on stochastically forced dynamo simulations on the firmed ground (see top panel of Fig.~5).

The presence of stochastic fluctuation breaks hemispheric coupling thus introduces a continuous phase shift in toroidal field evolutions between the two hemispheres, resulting in a change in parity. Interestingly we find that cyclic magnetic activity in the stronger hemisphere proceeds faster (i.e., complete more number of cycles in a given period) than the weaker hemisphere, which eventually flips the parity as time progresses. The top and bottom panel of Fig.~6 shows how the cyclic magnetic activity in the northern hemisphere shifts compared to the southern hemisphere with time in case of our dynamo simulation with stochastic fluctuation. We also find a significant negative correlation (Spearman correlation coefficient -0.64 with 99 \% confidence level) exists between the periodicity of the n$^{th}$ cycle and the amplitude of the (n+1)$^{th}$ cycle (Fig.~7). However, there exist no correlation between the periodicity of the n$^{th}$ cycle and the amplitude of n$^{th}$, (n-1)$^{th}$ and (n+2)$^{th}$ cycle (see Fig.~7). Thus in principle, one may use cycle length to predict the amplitude of the next solar cycle. This result also implies that solar dynamo has a memory of one cycle, which is in agreement with earlier observational results by \cite{sola02}. In summary, cycle length controls the amplitude of the next cycle. Considering above findings, one can say if solar activity in a certain hemisphere strongly dominates over the other hemisphere for several solar cycles, then it is more likely that magnetic activity in the stronger hemisphere will complete more number of cycles compared to the weaker hemisphere, eventually results hemispheric decoupling. 

We perform several numerical simulations by introducing different levels of fluctuations in both poloidal field sources and find that our model results are robust.

\section{Conclusions}

In order to figure out a qualitative relationship between parity and hemispheric asymmetry, we first decompose the solar surface magnetic field in terms of axial dipolar and quadrupolar moments. We find that hemispheric asymmetry significantly affects the nonlinear coupling between dipolar and quadrupolar modes of the solar magnetic field across the hemisphere and changes the parity over long time scales. We verify this result by performing kinematic solar dynamo simulations. We perform solar dynamo simulations where the poloidal field generation takes place through the combined effect of both the Babcock-Leighton mechanism and mean field $\alpha$-effect. By introducing stochastic fluctuations in the poloidal field source terms, we find dynamo solutions with changing parity. Earlier results in a different context (without any consideration of stochastic fluctuations in the dynamo source terms) has indicated that the parity issue may be related to the coupling between hemispheres \citep{chat06}. We demonstrate that the presence of stochastic fluctuations makes hemispheric coupling weak. Thus there may be a possible relationship between hemispheric asymmetry and parity change. The interplay of dipolar and quadrupolar modes can be interpreted as continuous nonlinear interactions between poloidal and toroidal components of the solar magnetic fields. An investigation reveals that parity changes are likely to occur only when one hemisphere strongly dominates over the other hemisphere for a long period persisting over several solar cycles. Our findings may open pathways for predicting parity flip in the Sun.

Systematic observations over the past century indicate that the solar magnetic field has always been in the dipolar parity state. However, it has been noted that there was large asymmetry in activity in the recovery phase of the Maunder minimum, wherein, the appearance of sunspots was almost confined to the southern hemisphere \citep{ribe93}. At this point, it is unclear whether this was related to a possible parity change in the Sun before or after the Maunder minimum. Independent simulations using low order dynamo models also predict the possibility of parity flips in the Sun \citep{beer98, knob98}. Thus, our results, taken together with other investigations point out that hemispheric coupling, parity shifts and the occurrence of grand minima episodes may be related. These interrelationships need to be investigated further and may provide a pathway for predicting parity shifts and the onset of grand minima episodes. 

\section*{Acknowledgements}
We are grateful to the Ministry of Human Resource Development, Council for Scientific and Industrial Research and University Grants Commission of the Government of India for supporting this research. We thank Prantika Bhowmik and Mayukh Panja for reading the manuscript and providing useful suggestions.












\end{document}